# SEARCH FOR RELIC NEUTRINOS AND SUPERNOVA BURSTS*

DAVID B. CLINE

*University of California Los Angeles, Department of Physics and Astronomy*
*Box 951547, Los Angeles, California 90095-1547   USA*
E-mail:  dcline@physics.ucla.edu

ABSTRACT

We describe the current situation concerning methods to search for relic neutrinos from the Big Bang and from all past supernovae (SNs). The most promising method for Big Bang neutrinos is by the interaction of ultra-high-energy (UHE) neutrinos. For supernova neutrinos, both Super Kamiokande- and ICARUS-type detectors will be important to study both $\bar{\nu}_e$ and $\nu_e$ fluxes. We also discuss a dedicated supernova burst observatory (OMNIS) being planned for three sites in the world. We also describe the possible analysis of the supernova type-II (SNII) neutrinos, including flavor mixing, that might be carried out in the future.

## 1. Introduction

A light neutrino mass between 1 eV and 100 eV would be highly significant for cosmology. In fact, if a neutrino contributes a fraction $\Omega_\nu$ of the closure density of the Universe, it must have a mass $m_\nu \approx 92\, \Omega_\nu\, h^2$ eV, where $h$ is the Hubble parameter in units of 100 km s$^{-1}$ Mpc$^{-1}$. Reasonable ranges for $\Omega_\nu$ and $h$ then give 1 eV to 30 eV as a cosmologically significant range. A neutrino with a mass in the higher end of this range (*i.e.*, $10 \leq m_\nu \leq 30$ eV) could contribute significantly to the closure density of the Universe. The cosmic background explorer (COBE) observation of anistropy in the microwave background, combined with observations at smaller scales, and the distribution of galaxy streaming velocities, have been interpreted as implying that there are two components of dark matter: hot ($\Omega_{HDM}$ ~ 0.3) and cold ($\Omega_{CDM}$ ~ 0.6). The hot dark matter (HDM) component could be provided by a neutrino with a mass of about 7 eV.[1-3] This shows the possible importance of astrophysical neutrinos. Figure 1 gives an overview of the different regions of diffuse neutrinos.

Despite the cosmological and particle physics interest in massive neutrinos, there are very few terrestrial experimental means for measuring the neutrino masses. The electron neutrino mass is constrained by the tritium end-point experiments to be less than about 3 eV. It is conceivable that $\nu_{\mu,\tau}$ accelerator neutrino-oscillation experiments, such as the NOMAD and CHORUS experiments at CERN, could be used to imply a mass in the cosmologically interesting range for $\nu_\mu$ or $\nu_\tau$. This would depend on there being a fairly large vacuum mixing between these neutrino flavors. Nucleosynthesis from supernovas could possibly provide a signature for neutrinos with these masses.

---

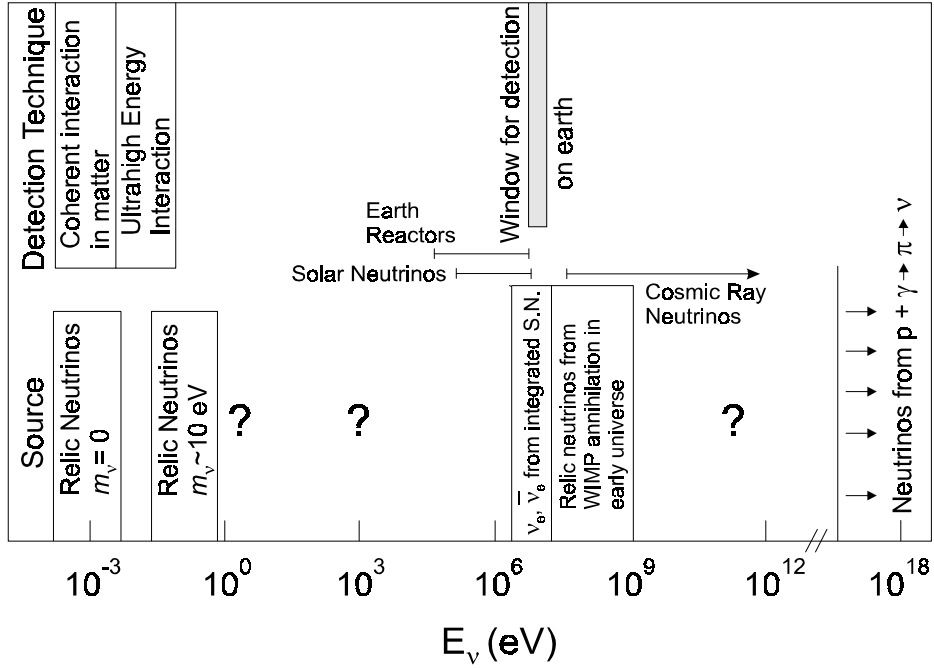

Fig. 1. Search for cosmic diffuse neutrinos.

## 2. Measuring the Neutrino Mass by Time of Flight

Perhaps the most straightforward and obvious nature of a massive neutrino would come from the lengthening in flight time from a distant supernova. For example, the flight time difference between $\nu_\tau$ and $\nu_e$ ($\bar\nu_e$) in seconds is

$$\Delta t = 5.14 \times 10^{-2} R_{\text{kpc}} \left[ \left( \frac{m_{\nu_e}}{E_{\nu_e}} \right)^{-2} - \left( \frac{m_{\nu_x}}{E_{\nu_x}} \right)^{-2} \right] \text{s} \quad ,$$

where $E_{\nu_x}$ is the neutrino energy in MeV, $m_{\nu_x}$ is in eV, and $R_{\text{kpc}}$ is the distance to the supernova in units of 10 kpc. A finite neutrino mass would alter the neutrino spectra in characteristic ways that could result in broadening and flattening of the observed signal.

Thus, neutrino masses might be obtained by comparing the observed neutrino signal with the signal expected from supernova models. Since detectors such as Super Kamiokande (SK) are relatively insensitive to $\nu_\mu$ and $\nu_\tau$, they are unlikely to measure cosmologically significant neutrino masses for these flavors. One of the neutral-current-based detectors being built at present is the Sudbury Neutrino Observatory (SNO). In order to measure the properties of neutrinos from supernovae, it is necessary to detect all of the neutrino flavors and the different processes are given in Table 1. In order to detect SNIIs, the detectors should record a number of events and operate like a supernova observatory, as illustrated in Table 2.

Table 1. Detection of $\nu_\mu$ and $\nu_\tau$ from supernova neutrinos in real time.

Two possibilities:

1. $\nu_x + e^- \to \nu_x + e^-$: Rate low because $\sigma_{\nu_x e}$ small; background from $\nu_x e \to \nu_e e$
2. $\nu_x + N \to \nu_x + N'^{4,5}$

   $N =$ D, C, O, NaCl, Pb, Fe ...

$$N' = n + X \begin{cases} \text{SNO} \\ \text{SNBO/OMNIS} \end{cases} \quad , \quad N' = \gamma + X \begin{cases} \text{SK} \\ \text{LVD} \end{cases}$$

SIGNAL DEPENDS ON $\nu_\mu, \nu_\tau$ ENERGY SPECTRUM

Table 2. Requirements of a supernova observatory.

- Life of Observatory $\geq$ rate (yr) for SNII on Milky Way Galaxy $\geq$ 20 – 40 yr
- Event Rate: $\sim 5 - 10$ K  $\bar{\nu}_e + P \to e^+ + n$
  
  $\sim$ Few K  $\nu_x + N \to \nu_x + N$
  
  $\nu_x = \nu_\mu + \nu_\tau$

  To: ∘ Fit model of SNII process
  
  ∘ Extract a neutrino mass or neutrino oscillation
  
  ∘ Learn about SNII explosion process

### 3. The Different Types of Supernova Neutrino Detectors and Neutrino Masses

Recently there has been real progress in supernova simulations giving an explosion.[4] These calculations give interesting predictions for the neutrino spectra. Detectors like the SK and SNBO may be able to detect such effects, however the SNBO detector may be of crucial importance for this study. Using these various detectors, it should be possible to detect a finite neutrino mass.[5] The characteristics of this detector are listed in Table 3.

In this analysis, we have assumed the existence of a very massive neutral-current detector (the SNBO), which we discuss next. By using these different detectors it will be possible to measure the $\mu$ or $\tau$ neutrino masses, which could determine a mass to $\sim 10$ eV. To go to lower mass, we need to use the possible fine structure in the burst; we have shown that it may be possible to reach $\sim 3$ eV with very large detectors in this case.[4] The detection of two-neutron final states would be useful for Pb detection.

Table 3. Properties of the (proposed) OMNIS/SNBO detector.

| | |
|---|---|
| Targets: | NaCℓ (WIPP site) |
| | Fe and Pb (Soudan and Boulby sites) |
| Mass of Detectors: | WIPP site ≥ 200 ton |
| | Soudan/Boulby sites ≥ 200 ton |
| Types of Detectors: | Gd in liquid scintillator |
| | $^6$Li loaded in the plastic scintillators that are read out by scintillating-fiber–PMT system |

## 4. The Proposed Supernova Burst Observatory (SNBO)

The major problem of supernova detection is the uncertain period of time between such processes in this Galaxy. In addition, complimentary detectors should be active when the supernova goes off in order to gain the maximum amount of information possible about the explosion process and neutrino properties. In Table 2, we list some of the requirements of such an ideal supernova observatory.

Lacking an ideal observatory, a group of us have been studying a very large detector, SNBO.[5] Table 3 gives some of the guidelines for this detector,[6] and Fig. 2 gives additional detailed simulation results in the development of the SNBO detector.[7] We have located a possible site for the observatory near Carlsbad, NM, which is the WIPP site (shown in Fig. 3). We have studied the radioactive background at this site (measured by the OSU–UCLA group) and find it acceptable for a galactic supernova detector. We find less than one neutron per hour detected in a 6-ft $BF_3$ counter. This leads to the expectation that the background for a galactic supernova is much smaller than the signal at this site. In Fig. 4, we give the expected event rate for various channels in the various detectors around the world.

## 5. Search for Relic Neutrinos

Two general methods have been suggested to detect relic neutrinos (see Table 4):
1. By the coherent interaction force given to a sensitive system (like a torsion balance), which will depend on the neutronic mass, or
2. By interaction of UHE neutrinos with the relic neutrinos to produce a $Z^o$, for example, which will also likely require a small relic neutrino mass.

In Fig. 5, we show the cross section for the neutrinos with matter and the coherence length as a function of neutrino energy. There is a clear dependence on neutrino mass.[8] Recently

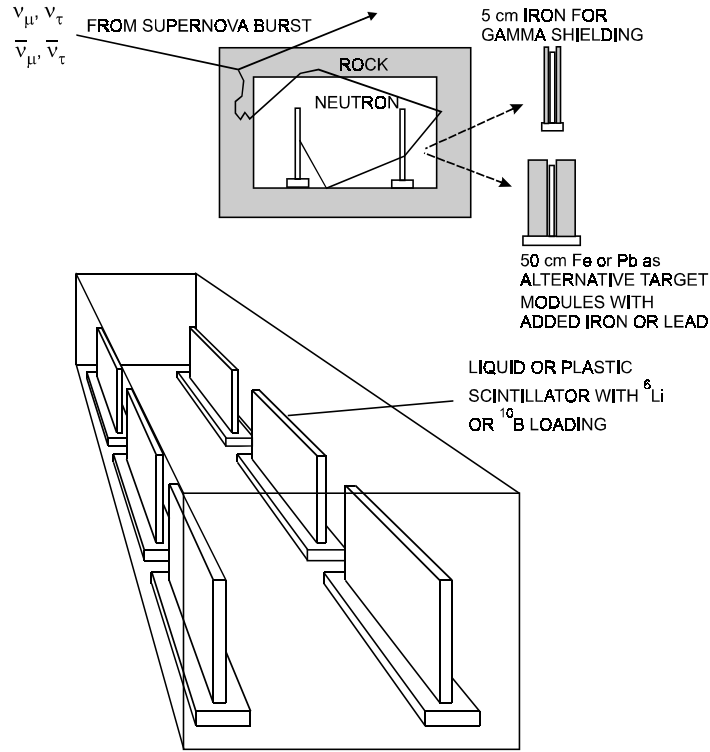

Fig. 2. Illustrative arrangement of detector modules along rock tunnel; also shown is the use of supplementary iron or lead for gamma shielding or as an alternative neutrino target.

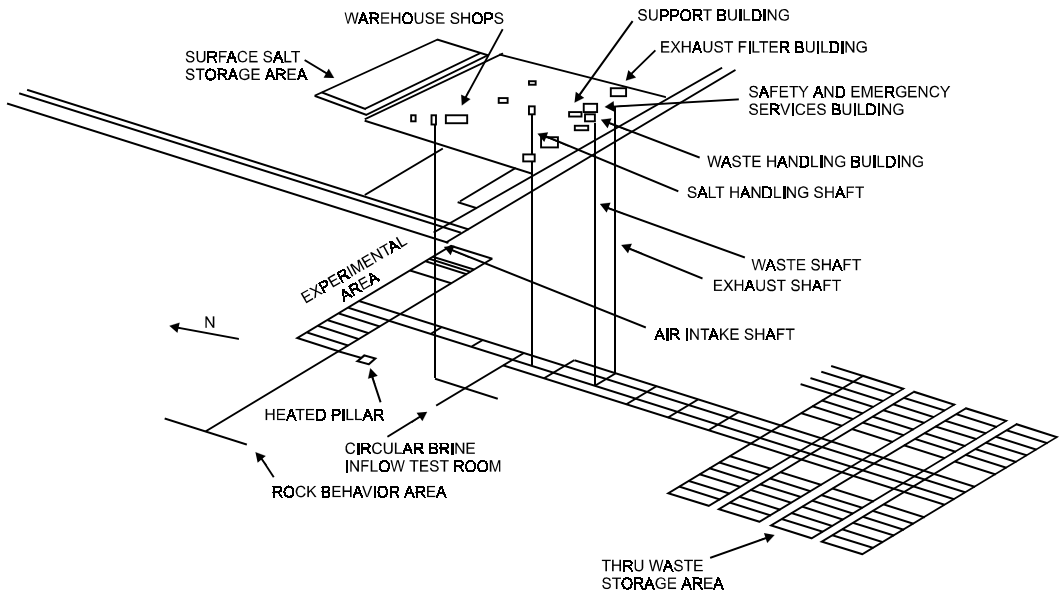

Fig. 3. Isometric view of the surface and underground (looking toward the Northeast) of the WIPP site near Carlsbad, NM.

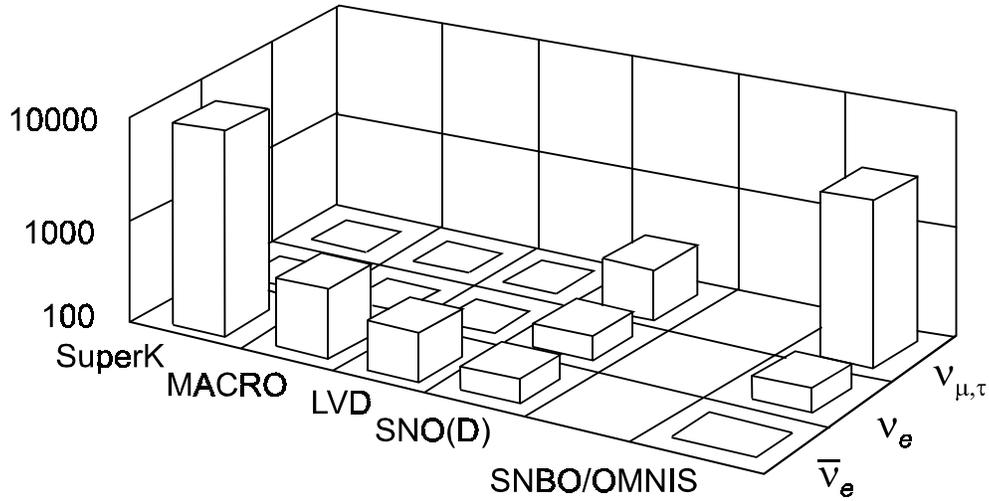

Fig. 4. Comparison of world detectors (event numbers for supernovae at 8 kpc).

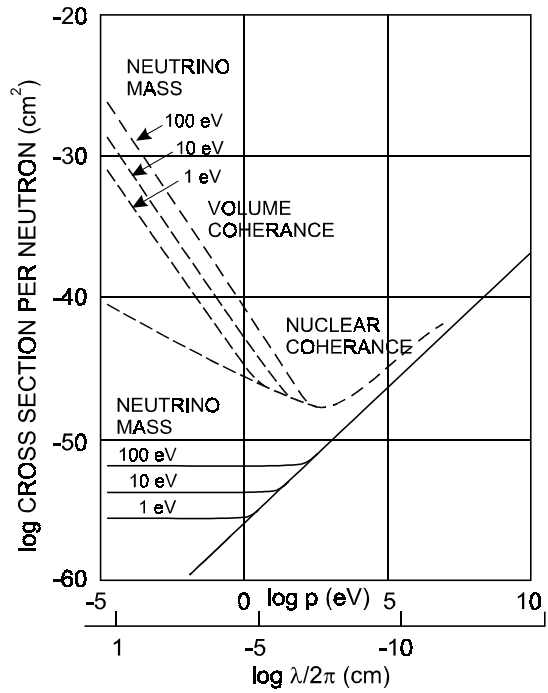

Fig. 5. Neutrino elastic scattering cross section vs momentum $p$ and corresponding wavelength $\lambda$ (—, single particle scattering; —, coherent scattering). [P. F. Smith, private communication (1999)]

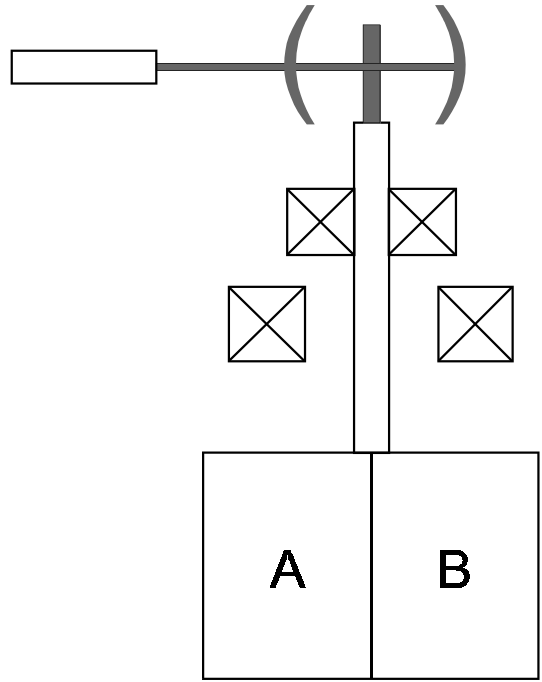

Fig. 6. Schematic of the torsion oscillator. The target consists of two hemi-cylindrical masses with similar densities but different neutrino cross sections. The mass is suspended by a "magnetic hook" consisting of a superconducting magnet in persistent mode floating above a stationary magnet. The rotation angle is read out with a tunable optical cavity and an ultra-stable laser.

Table 4. Techniques to search for relic neutrinos.

## OPTIONS FOR RELIC NEUTRINOS – EXAMPLES

1. **STANDARD MODEL**

   $m_{\nu_{e,\mu,\tau}} \simeq 0$ , $T_\nu \simeq 1.9°K$ , $\mu_\nu \simeq 10^{-18} (m_\nu) \mu_B$ ,

   $P_\nu \simeq (5 - 2) \times 10^{-4}$ eV/$c$ , $\alpha \nu_e = 3 \times 10^{12}$ cm$^2$ s

   $$\boxed{n_\nu = \frac{3}{4} n_\gamma}$$

2. **MASSIVE NEUTRINOS**

   $m_{\nu_{\mu,\tau}} \simeq 30$ eV (either $\mu$ or $\tau$; $m_{\nu_e} = 11$ eV)

   (A) Unclustered in Galaxy    (B) Clustered in Galaxy

3. **NEUTRINOS WITH UNUSUAL PROPERTIES**

   $\mu_\nu \simeq 10^{-10} \mu_B$ , $\mu_\nu \simeq ? (30$ eV) , (Transition $\nu_B$ or $\nu_\mu$)

   (i.e., clustered and with a large magnetic moment)

4. **(RELIC NEUTRINOS) FROM THE DECAY OF MASSIVE RELIC NEUTRINOS**

   Example: $M_{\nu_\tau} = 1$ MeV

   Assume dominant decays:

   $\nu_\tau \rightarrow \nu_\mu + \phi$ if $\tau_{\nu_\tau} \gtrsim 10^{16}$ s

   $\rightarrow 3\nu$ , (Assume $m_{\nu_\mu} \sim 0$)

   $E_{\nu_\mu} \approx \dfrac{1/2 \text{ MeV}}{1 + z}$ , Flux

   MORAL:
   RELIC NEUTRINOS HAVE NOT BEEN DETECTED;
   HENCE WE DO NOT KNOW THE SPECTRUM

C. Hagman has suggested that a very sensitive to torsion balance may someday be sensitive enough (shown in Fig. 6). Although a recent meeting at UCLA covered this subject, there are many issues that are unresolved.[8]

The other method that uses UHE neutrinos is illustrated in Fig. 7A, and was just proposed by T. Weiler.[9] The detection of the UHE neutrinos could be carried out as is shown in Fig. 7B.

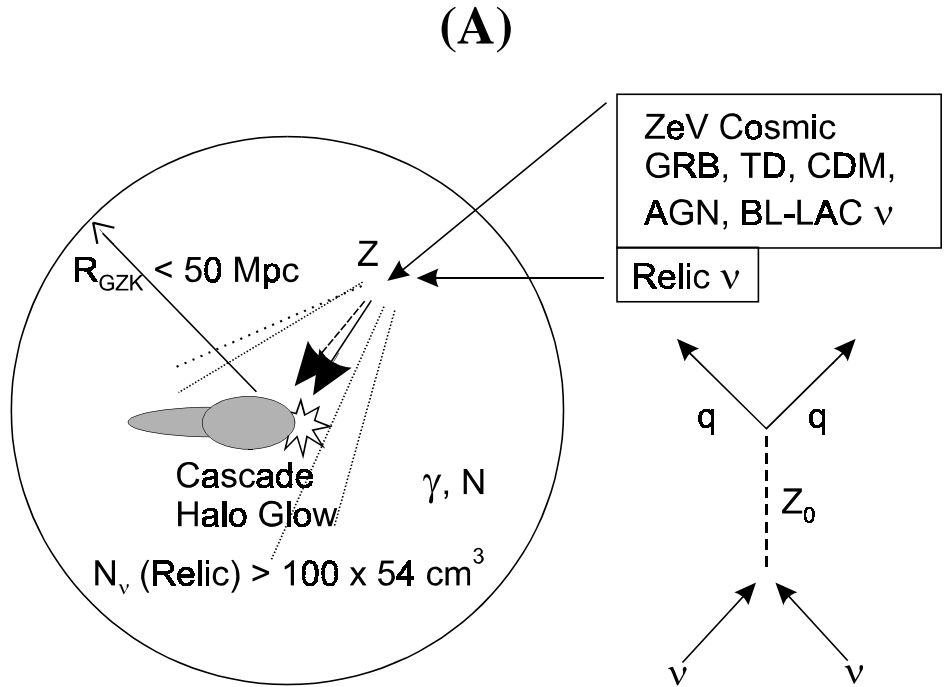
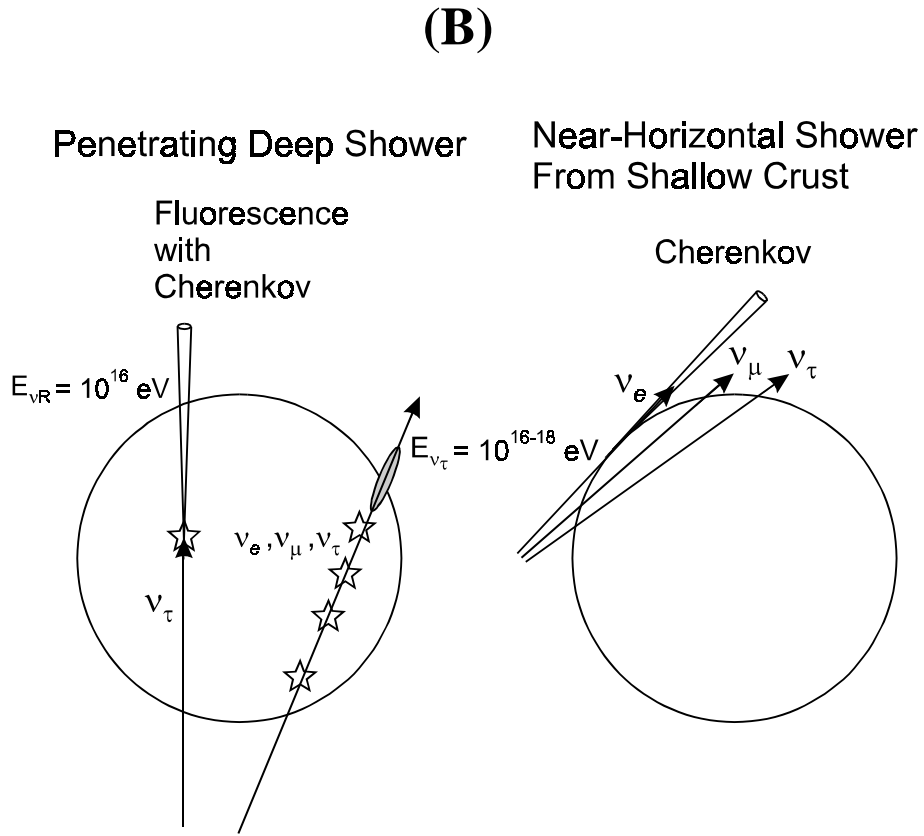

Fig. 7. (A) "ν-tomography" of hot dark matter of Virgo cluster with cosmological neutrinos.[9]
(B) Illustration of backward showers initiated by earth-penetrating cosmological neutrinos.

## 6. Search for the Integrated Flux of Supernova Neutrinos

Another kind of relic neutrino that which arises from the integrated flux from all past SNIIs. Figure 8 shows a schematic of these (and other) fluxes. These fluxes could be modified by transmission through the SNII environment, as discussed by G. Fuller and others and as shown in Fig. 9.

The detection of $\bar{\nu}_e$ from the relic supernovae may someday be accomplished by the SK detector; it would be as interesting to detect $\nu_e$ with an ICARUS detector, as illustrated in Table 5. High-energy $\nu_e$ would come from $\nu_{\mu,\tau} \rightarrow \nu_e$ in the supernova. A window of detection occurs between the upper solar neutrino energy and the atmospheric neutrinos, as first proposed by D. Cline and reported in the first ICARUS proposal (1983-1985). The ideal detector to observe this is a large ICARUS liquid-argon detector. The detector and method are illustrated in Figs. 10 and 11.

## Acknowledgments

We wish to thank the ICARUS team and G. McLaughlin, G. Fuller, and W. Haxton.

Table 5. Detection of $\nu/\bar{\nu}_e$ relic neutrino flux from time integrated SNIIs.

1. Relic $\nu/\bar{\nu}_e$ from all SNIIs back to $Z \sim 5$: $<E_\nu> \sim 1/(1 + Z)<E_\nu>$.
2. Detection would give integrated SNII rate from Universe
   – Window of detection [D. Cline, ICARUS proposal, 1984].
3. Neutrino oscillations in SNII would give $\nu_x \rightarrow \nu_e$ with higher energy than $\bar{\nu}_e$.
4. Detect $\bar{\nu}_e$ with SK or ICARUS. Attempt to detect $\nu_x/\nu_e$ detection.

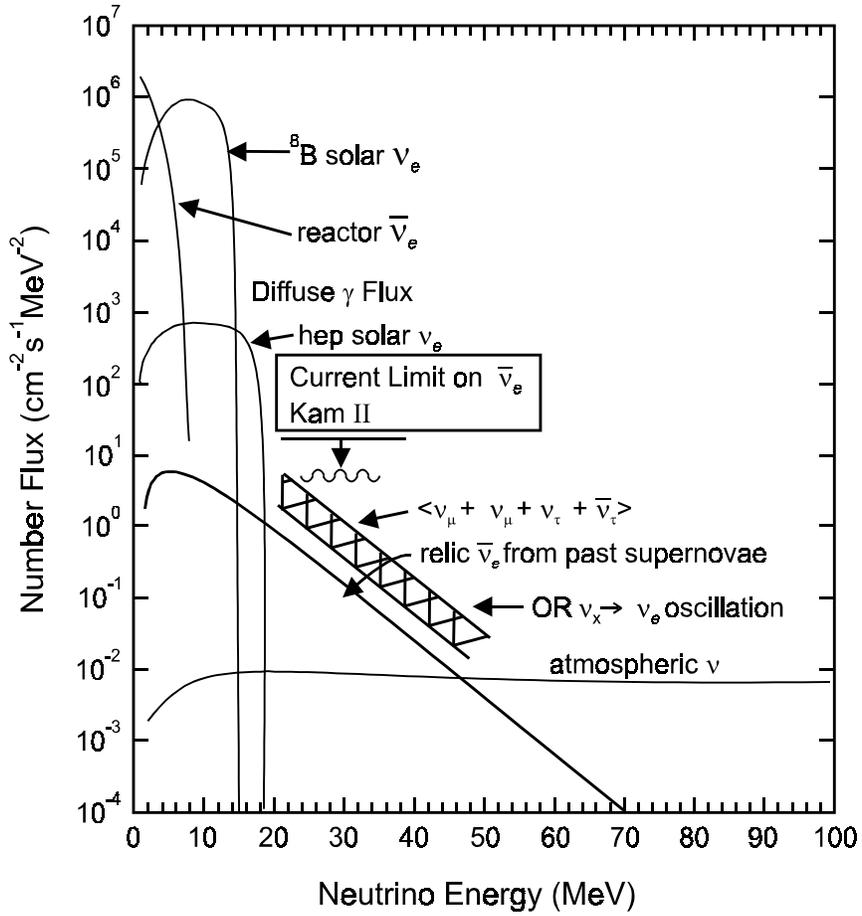

Fig. 8. Relic neutrinos from past supernova. Note: $v_x \to v_e$ in the supernova can boost the energy of the $v_e$ if we find $<E v_e> >> <E\bar{v}_e>$. This will be a signal for neutrino oscillation in supernovae! and measure $\sin^2 \theta x_e$.

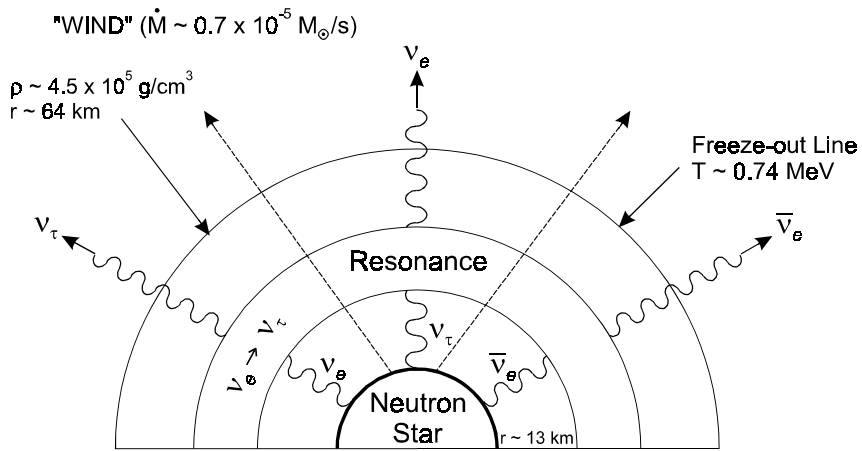

Fig. 9. Transmission of various neutrinos through the outer part of the SNII. It is possible that $v_\tau \to v_e$ conversion can occur in this environment. [G. Fuller, private communication (1999)]

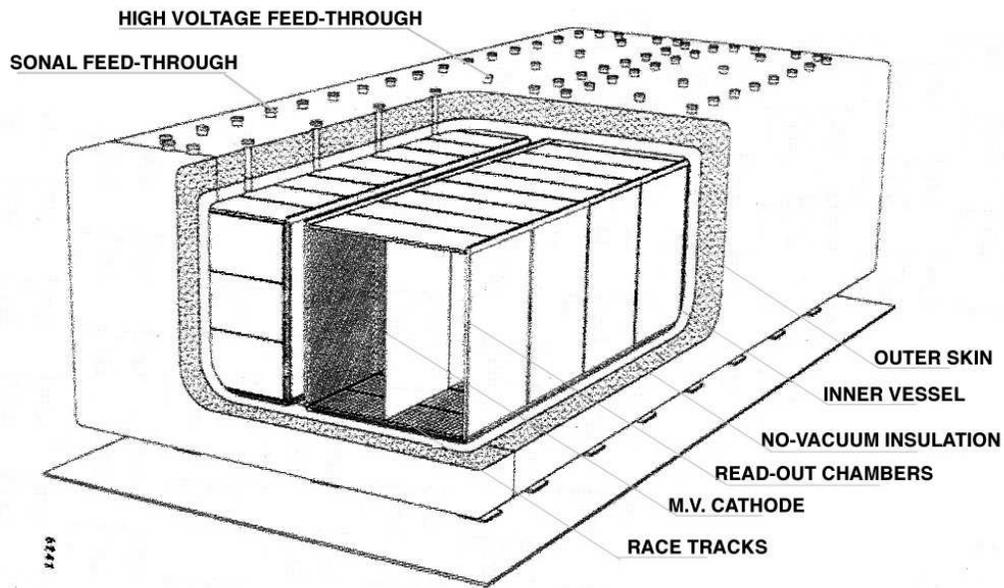
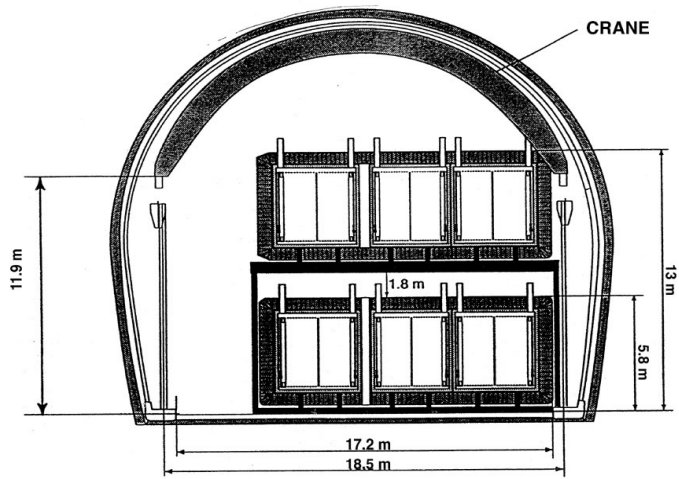
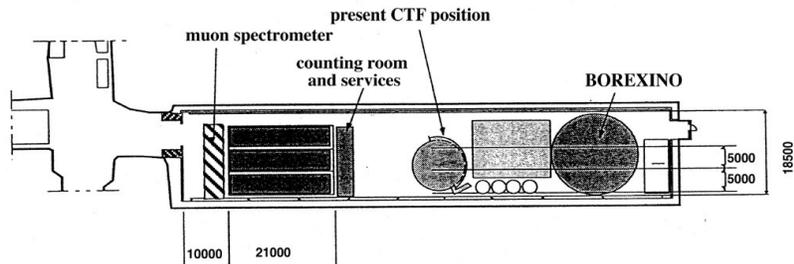

Fig. 10. Drawings of the ICARUS 600-ton module (top) and of the 1st 600-ton module in Hall C (bottom).

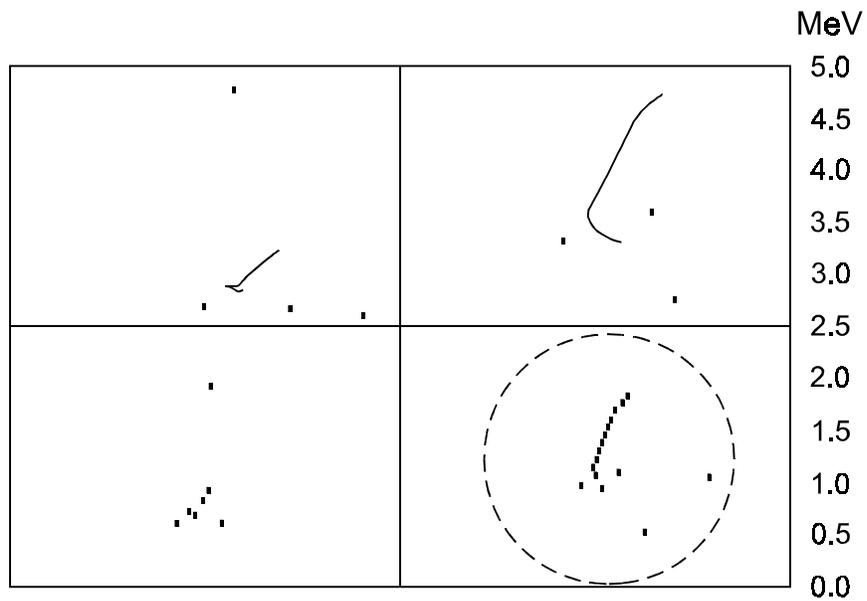

Fig. 11. Absorption event as generated by the GEANT Monte Carlo program in two wire planes put at 60 deg. In the bottom, the same event is shown after digitization. The gray scale of each pixel is proportional to the deposited charge. The resolution in the horizontal axis (drift direction) is 0.1 mm, and in the vertical axis it is 3 mm (wire pitch). The projected track length is ~ 3 cm, the main electron energy is 7 MeV, the associated energy is 2 MeV, and the associated multiplicity is 3.